\documentclass[prb,showpacs,twocolumn]{revtex4}
\usepackage{amsmath,graphicx,amssymb}

\begin{document}

\title{Chain-induced effects in the Faraday instability on ferrofluids \protect\\ in a horizontal magnetic field}
\author{V. V. Mekhonoshin}
\email[]{vladik@mpipks-dresden.mpg.de}
\author{Adrian Lange}
\email[]{adlange@mpipks-dresden.mpg.de}
\affiliation{Max-Planck-Institut f\"ur Physik komplexer Systeme, N\"othnitzer Str. 38, D-01187, Dresden, Germany}

\date{\today}

\begin{abstract}
{ The linear stability analysis of the Faraday instability on a
viscous ferrofluid in a horizontal magnetic field is performed.
Strong dipole-dipole interactions lead to the formation of chains
elongated in the field direction. The formation of chains results
in a qualitative new behaviour of the ferrofluid. This new behaviour
is characterized by a neutral stability curve similar to that observed
earlier for Maxwell viscoelastic liquids and causes a significant
weakening of the energy dissipation at high frequencies. In the
case of a ferrofluid with chains in a horizontal magnetic field,
the effective viscosity is anisotropic and depends on the field
strength as well as on the wave frequency.}
\end{abstract}

\maketitle

\section{\label{intro} Introduction}

Magnetic fluids (or ferrofluids) are colloidal dispersions of single domain
nanoparticles in a carrier liquid. The fascination of ferrofluids
stems from the combination of a normal liquid behaviour with the sensitivity
to magnetic fields. This enables the use of magnetic fields
to control the flow  of the fluid, giving rise to a great variety of new phenomena
and to numerous technical applications
\cite{Ferrohydrodynamics}. One of the peculiarities of ferrofluids is the mutual
influence of the microstructure and the rheological properties of the fluid.
Due to strong interparticle
interactions various aggregates can be formed in a ferrofluid
\cite{deGen,Jordan,Taketomi,jcis}. On the one hand, the formation
of the aggregates changes the effective viscosity of the fluid. On
the other hand, the motion of the fluid influences the structure
of the aggregates.

In an applied magnetic field chains containing
several particles are the favored form of an aggregate
and are formed by the dipolar interactions. There is
a similarity between the chains of dipoles and the macromolecular
chains in polymer solutions. In both systems a network of chains is
coupled with a viscous carrier liquid. This viscous coupling implies,
that the relaxation character of the chain dynamics leads to a viscoelastic
behaviour of the solution. The viscoelastic behaviour is reflected in a
dependence of the stress tensor on the history of the system.
The theoretical treatment of the viscoelasticity of polymer solutions is
usually based on a phenomenological model. There are three well known models
for an isotropic linear viscoelastic liquid. In the Maxwell model a
viscoelastic element is a combination of a purely elastic spring with a
purely viscous dashpot. In the Bingham two-component model these elements
are connected parallel to each other, and the Jeffreys element is a
superposition of the two previous ones. A comprehensive review
of the theoretical basics and the existing models was done by Bird in
Refs.\ \onlinecite{Bird,Bird1}. An overview of the theoretical approaches
to the polymer dynamics is given  by Doi and Edwards \cite{Doi}.

A ferrofluid with chains in a magnetic field is an anisotropic system. This
makes it similar to a nematic liquid crystal (NLC), whose rheology can be
described by a group of models usually referred as Ericksen-Leslie-Parody
(ELP) models.  The stress tensor in Refs.\ \onlinecite{Ericksen,Ericksen1,Leslie,Parodi}
involves five independent viscosity coefficients, which are scalar functions
of the density and the temperature. ELP models are widely used in the hydrodynamics
of NLC. These models were employed in the analysis of a flow instability
\cite{Leslie1} and in studying periodic patterns \cite{Simoes}. In these papers
as well as in Refs.\ \onlinecite{Leslie2,Clark} the influence of an external field
was taken into account. Readers who are interested in a deeper insight into ELP
models are referred to the books of de Gennes \cite{deGennes_book} and
Chandrasekhar \cite{Chandrasekhar_book}.

Since the susceptibility of ferrofluids is much higher than that of liquid crystals,
the effects caused by external field should be more pronounced. The
theoretical analysis of the rheology of a colloidal suspension containing ellipsoidal
particles in a field was performed by Pokrovskij in Refs.\
\onlinecite{Pokr_art,Pokr_book}. The model gives a constitutive
equation of the ELP kind, where the viscosity coefficients are expressed in terms
of the parameters of the suspension and the applied field. Electro- and magnetorheological
fluids present a group of suspensions, whose behaviour is close to that of
ferrofluids. The viscoelastic properties of electro- end magnetorheological
fluids were studied experimentally in Refs.\
\onlinecite{Hanaoka,Yuxian,Bossis}. Particularly
it was shown that external fields can influence the rheological
response of the fluid by changing the complex shear modulus and yield stress.
Zubarev and Iskakova \cite{Zubarev}
used the results of Refs.\ \onlinecite{Pokr_art,Pokr_book} to
obtain a constitutive equation for a ferrofluid with chains for
the case of a weak flow of any kind. In the present paper the
model suggested in Ref.\ \onlinecite{Zubarev} is applied to the
linear stability analysis of the Faraday instability.

The Faraday instability denotes the parametric generation of
standing waves on the free surface of a fluid subjected to
vertical vibrations. The study of this phenomenon dates back to
the observations by Faraday \cite{Faraday} in 1831. The initially
flat free surface of the fluid becomes unstable at a certain
intensity of the vertical vibrations of the whole system. As a
result of the instability, a pattern of standing waves is formed
at the fluid surface. The typical response is subharmonic, i.e.,
the wave frequency is half the frequency of the excitation. A
harmonic response can be observed on a shallow fluid at low
frequencies \cite{Muller1}. Faraday waves allows one to
investigate symmetry breaking phenomena in a spatially extended
nonlinear system. Therefore they experience a renewed interest in
recent years. Detailed experimental studies of the various
patterns on a viscous fluid have been performed \cite{Fauve,Muller93,Kudrolli,Arbell1,Arbell},
where a one-frequency
as well as a two-frequency forcing were applied. Among the
observed patters are parallel rolls \cite{Fauve}, hexagons \cite{Fauve},
a twelvefold quasi-pattern \cite{Fauve},
triangles \cite{Muller93}, superlattices formed by small and large
hexagons \cite{Kudrolli}, squares
\cite{Muller93,Kudrolli,Arbell1,Arbell}, and rhomboid pattern
\cite{Arbell}.

The comprehensive linear stability analysis of the Faraday
instability on an arbitrarily deep layer of a viscous non-magnetic
fluid has been performed by Kumar and Tuckerman \cite{Kumar}. This
analysis was tested experimentally \cite{Bechhoefer} and  an
excellent agreement between the  predicted and experimental data
was found. In Refs.\ \onlinecite{Muller1,Cerda} the low frequency
region is studied particularly. Bicritical points, where
transitions from one type of response to others occur, are
predicted and experimentally confirmed \cite{Muller1}. In Ref.\
\onlinecite{Tuckerman,Kumar1} an analogy between the Faraday
instability and a periodically driven version of the
Rayleigh-Taylor instability is exploited. Based on that analogy
in Ref.\ \onlinecite{Tuckerman} a scaling law is suggested, which
satisfactorily describes the behaviour of the system in a wide
range of parameters. S. Kumar \cite{Kumar1} discusses the
mechanism of the wave number selection in the Faraday instability
on high-viscous fluids.

In our previous paper \cite{pre02}, the Faraday instability on a
chain-free ferrofluid was studied. A nonmonotonic dependence of
the stability threshold on the magnetic field is found at high
frequencies of the vibrations. It was revealed that the magnetic
field can be used to select the first unstable pattern of Faraday
waves. In particular, a rhombic pattern as a superposition of two
different oblique rolls can occur.

The Faraday instability of a viscoelastic non-magnetic liquid was
studied experimentally in Ref.\ \onlinecite{Mueller_W}, where a harmonic
response was detected. In Refs.\ \onlinecite{Muller_ve,S_Kumar1,S_Kumar}
the Maxwell model of viscoelastic liquid was used in the theoretical analysis.
The authors observed pronounced changes in the neutral stability curves.
Particularly, the tongues related to the harmonic response became abnormal.
Such a tongue has no tip and all tongues of higher order
are inside this abnormal tongue.

The aim of the present paper is to investigate the role of the
chains in the Faraday instability on a ferrofluid in a horizontal
magnetic field. The formation of the chains leads to a dramatic
increase of the magnetization  relaxation time, changes the
effective viscosity of the suspension, and increases the
susceptibility of the ferrofluid. A number of model ferrofluids
are investigated in a wide range of the parameters of the system
to study the relative importance of those effects.

\section{\label{system}System and basic equations}
\subsection{\label{fluid}Model}

A dielectric, viscous, and incompressible magnetic fluid with
constant density $\rho$ is considered, which contains particles of
the equal size. The strength of the dipole-dipole interactions is
characterized by the coupling constant
$\varepsilon=\mu_0m^2/(16\pi R^3 k_B T)$, which is the ratio of the
energy of interaction between two particles at the minimal
separation with head-to-tail orientation of their magnetic moments
to the thermal energy. Here $m=4M_s\pi R_0^3/3$ is the magnetic
moment of a particle, $R=R_0+\delta$ is the hydrodynamic radius of
the particle, $M_s$ is the magnetization of the magnetic material,
$R_0$ is the radius of the magnetic core of a particle, $\delta$
is the thickness of the nonmagnetic layer, $k_B$ is the Boltzmann
constant, and $T$ is the temperature. The interaction of a
particle with the applied field $H$ is measured by the Langevin
parameter $\kappa=\mu_0mH/(k_BT)$. In the case of a magnetic fluid
with low particle volume fraction $\varphi$, interactions between
chains can be neglected. Assuming that the chains are straight and
rigid, Zubarev and Iskakova determined the size distribution of
chains, which minimizes the free energy of such a magnetic fluid
and is given by \cite{Zubarev}:

\begin{equation}
g_n=\frac{x^n}v \frac{\sinh(\kappa n)}{\kappa n}\exp(-\varepsilon),
\label{sizes}
\end{equation}

$$
\begin{aligned}
x=&\left[2y\cosh\kappa+\sinh\kappa\phantom{\sqrt{x^2}}\right.\\
 &\; \left.-\sqrt{(2y\cosh\kappa+\sinh\kappa)^2-4y^2}\right]/(2y),
\end{aligned}
$$

\noindent where $v$ is the volume of a particle and
$y=\kappa\varphi\exp(\varepsilon)$. Each chain containing $n$
particles is modelled by an uniaxial ellipsoid with semi-axes
equal to $nR$ and $R$. This keeps the solid phase volume density
unchanged and allows one to use the results of Pokrovskij
\cite{Pokr_art,Pokr_book}. The viscous stress tensor for a system
of uniaxial ellipsoids consists of a symmetric and antisymmetric
part

\begin{equation}
\sigma_{ik}=\sigma_{ik}^{(s)}+\sigma_{ik}^{\text{(as)}},
\label{v_stress}
\end{equation}

\noindent where

$$
\begin{aligned}
\sigma_{ik}^{(s)}=&2\eta\gamma_{ik}+\eta\\
&\times \biggl\langle\biggl\langle \biggl[ 2\alpha_n\gamma_{ik}-\rho_n\left\langle e_je_s\right\rangle_n\delta_{ik}\gamma_{js}\\
&+\left(\zeta_n+\beta_n\lambda_n\right)\left(\left\langle e_ie_j\right\rangle_n\gamma_{jk}+\left\langle e_ke_j\right\rangle_n\gamma_{ji}\right)\\
&+\beta_n\bigl(\Omega_{ij}\left\langle e_je_k\right\rangle_n+\Omega_{kj}\left\langle e_je_i\right\rangle_n\bigr)-\beta_n\frac d{dt}\left\langle e_ie_k\right\rangle_n\\
&+\left(\chi_n-2\lambda_n\beta_n\right)\left\langle e_ie_ke_je_s\right\rangle_n\gamma_{js}\biggr] \biggr\rangle\bigg\rangle,\\
\sigma_{ik}^\text{(as)}=&\frac{\kappa k_BT}{2v}\bigl\langle\bigl\langle \left\langle e_i\right\rangle_n-\left\langle e_k\right\rangle_nh_i\bigr\rangle\big\rangle.
\end{aligned}
$$

\noindent Both parts contain the geometric factors $\alpha_n$,
$\beta_n$, $\chi_n$, $\lambda_n$, $\rho_n$, and $\zeta_n$, which
are entirely determined by the aspect ratio of an ellipsoid
corresponding to a {\it n}-particle chain
\cite{Pokr_art,Pokr_book,Zubarev,correction}. $\eta$ is the
viscosity of the carrier liquid,
$\gamma_{ik}=(\partial_ku_i+\partial_iu_k)/2$ and
$\Omega_{ik}=(\partial_ku_i-\partial_iu_k)/2$ are the symmetric
and antisymmetric parts of the tensor of velocity gradients,
$\mathbf{e}$ and $\mathbf{h}=\mathbf{H}/H$ are the unit vectors
along the chain axis and the magnetic field, and the following
notations are used:

$$
\begin{aligned}
\langle\langle\cdots\rangle\rangle&=\sum_n\cdots n v g_n,\\
\langle\cdots\rangle_n&=\int{\cdots\mathbf{e}\psi_n(\mathbf{e})d\mathbf{e}},\qquad\langle\cdots\rangle_n^0=\int{\cdots\mathbf{e}\psi_n^0(\mathbf{e})d\mathbf{e}}.
\end{aligned}
$$

\noindent Here, $\psi_n^0(\mathbf{e})$ and $\psi_n(\mathbf{e})$
are the equilibrium and the nonequilibrium angular distribution
functions. The former is known due to the classical Langevin
model and the latter is the solution of the Fokker-Planck equation
(see Refs.\ \onlinecite{Pokr_art,Pokr_book,Zubarev}). The exact solution for
the case of an arbitrary field is unknown. Therefore, Zubarev and
Iskakova suggested an approximation for $\psi_n(\mathbf{e})$:

\begin{align}
\psi_n(\mathbf{e})=&\psi_n^0(\mathbf{e})\left[1+a_i\left(e_i-\left\langle e_i\right\rangle_n^0\right)\right.\nonumber\\&\left.+b_{ik}\left(e_ie_k-\left\langle e_ie_k\right\rangle_n^0\right)\right],
\label{orient}
\end{align}

\noindent where $\mathbf{a}$ and $\tensor{b}$ are a vector and a
symmetric tensor. They can be found from the equations for the
first $\langle e_k\rangle_n$ and the second $\langle
e_ie_k\rangle_n$ moments of $\psi_n(\mathbf{e})$, which are
derived from the Fokker-Planck equation
\cite{Pokr_art,Pokr_book,Zubarev} with the accuracy up to the
linear terms with respect to the velocity gradients. The moment
equations for $\langle e_k\rangle_n$ and $\langle e_ie_k\rangle_n$
involve two relaxation times, $\tau_1 = 1/(2D)$ and
$\tau_2=1/(6D)$, where $D\sim 1/\eta$ is the coefficient of rotational
diffusion.

One can easily see, that the magnetization of the ferrofluid is

\begin{equation}
\mathbf{M}=M_s\frac{R_0^3}{R^3}\bigl\langle\bigl\langle\:\langle \mathbf{e}\rangle_n\:\bigr\rangle\bigr\rangle.
\end{equation}

\noindent If $\mathbf{a}$ and $\tensor{b}$ are known one can
relate the viscous stress tensor $\tensor{\sigma}$ and the
magnetization to the tensor of velocity gradients for the given
value of magnetic field strength $\mathbf{H}$. The total stress tensor
$\tensor{T}$ reads

\begin{equation}
T_{ik}=-\left(p+\mu_0\int_0^H{MdH'}+\frac{\mu_0}2H^2\right)\delta_{ik}+H_iB_k+\sigma_{ik},
\end{equation}

\noindent where $\mathbf{B}=\mu_0\left(\mathbf{H}+\mathbf{M}\right)$, $p$ is the
pressure, and $\mathbf{B}$ the induction of the magnetic field.

\subsection{\label{Setup}The system}

The above model is used to analyze the stability of
the free surface of a ferrofluid in the following setup. The
laterally infinite ferrofluid layer of arbitrary depth $d$ is
subjected to a homogeneous dc horizontal magnetic field
\begin{figure}[tbp]
 \includegraphics[scale=0.45]{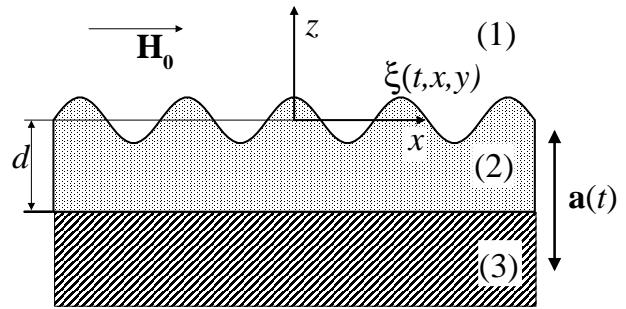}
\caption{A horizontally unbounded ferrofluid layer (2) is  placed
in a nonmagnetic container (3) with air (1) above. The system is
subjected to a horizontal magnetic field $\mathbf{H}_0$ and
harmonic vertical vibrations $\mathbf{a}(t)$.
}
  \label{setup}
\end{figure}
$\mathbf{H}_0=(H_0,0,0)$ and harmonic vertical vibrations (Fig.\
\ref{setup}). The plane $z=0$ coincides with the nondeformed
surface of the ferrofluid. The fluid layer is bounded from below
by the bottom of the nonmagnetic container and has a free surface
described by $\xi(t,x,y)$ with air above.

Due to zero electrical conductivity of the fluid, the static form
of the Maxwell equations is used for the magnetic field in all three
media. The fluid motion is governed by the continuity equation and the
conservation law of the linear momentum

\begin{subequations}
\label{gov}
\begin{align}
\text{div}\, \mathbf{u} &= 0\; ,\label{cont}\\
\rho\left[\frac{\partial u_i}{\partial t}+\left(\mathbf{u}\,\text{grad} \right)  u_i\right] &= \partial_jT_{ij}^{(2)}-\rho g(t)\delta_{i,3},\quad i=1,2,3.
\label{Navier}
\end{align}
\end{subequations}

\noindent  The vertical vibrations add a periodic  term to the
gravity acceleration $ \mathbf{g}_0$, i.e., a modulated value $
\mathbf{g}(t) = (0,0,-g_0-a\,\cos(\omega t))$ appears in the
equations of the fluid motion. Here $a$ is the acceleration
amplitude and $\omega$ is the angular frequency of the vibrations.
The governing set of equations has to be supplemented by the
boundary conditions, which are the same as in Ref.\ \onlinecite{pre02}.

\section{\label{Floquet}Linear stability analysis}

Following the standard procedure~\cite{Kumar,Muller}, the
governing equations and the boundary conditions have been
linearized in the vicinity of the nonperturbed state
\begin{align*}
\mathbf{u}&=0,\quad \xi=0,\quad \mathbf{H}^{(i)}=\mathbf{H}_0,\quad i=1, 2, 3 \; ,\\
p_0&=p^{(1)}-\frac{\mu_0}{2}M_0H_0-g(t)z \; ,
\end{align*}

\noindent where $M_0$ and $p_0$ are the magnetization and the
pressure in the unperturbed state. In order to construct the
linearized governing equations for the small perturbations, which
encompass the magnetic field strength $\mathbf{H}_1$, the pressure
$p_1$, and a nonzero velocity $\mathbf{u}$ of the fluid, it is
necessary to expand all quantities in Taylor series.

The above defined distribution functions $g_n$ and $\psi_n$ are
affected by the small perturbations. Since the formation and the dissociation
of chains are connected with the diffusion of particles in the suspension,
they are rather slow processes. It can be estimated by the
Schmidt number $Sc= \eta/(\rho D_0)$ which relates the characteristic time
for mass transport by flow to the characteristic time for mass transport by
diffusion. For a typical
ferrofluid the Schmidt number is about $5\cdot 10^6$ with
$\eta=0.1$ kg/(ms), $\rho = 1020$ kg/m$^3$, and the Brownian diffusion
coefficient $D_0=2\cdot 10^{-11}$ m$^2$/s $\;$\cite{voelker}.
Therefore we neglect the changes in $g_n$ caused
by the perturbations. This implies that the size distribution of chains
does not depend either on the spatial coordinates or on time. In Refs.\
\onlinecite{Pokr_art,Pokr_book,Zubarev} only spatially homogeneous
systems are considered. An extension to the case of a spatially
inhomogeneous system results in additional convective terms in the
dynamic equations for $\psi_n$ [Eq.\ (12) in Ref.\
\onlinecite{Zubarev}]. Since these terms are of the form
$\left(\mathbf{u}\, \text{grad}\right)\psi_{n,1}$, i.e., they are of
second order with respect to the perturbations, they can be
neglected in a linear stability analysis. As long as the typical
length scale of a chain ($\sim 10^{-8}\dots 10^{-7}$ m) is smaller than that of the spatial
variation of the perturbations (typically $\sim 10^{-3}\dots 10^{-2}$ m), Zubarev's model can be applied in
the present setup. Our results (see Sec.\ \ref{results})
suggest that this condition is always fulfilled.

The symmetry of the system possesses a number of restrictions on
$\mathbf{a}$ and $\tensor{b}$. The change of $y$ to $-y$ and $z$
to $-z$ should not change the components $a_x$, $b_{xx}$,
$b_{yy}$, and $b_{zz}$, whereas this transformation leads to the
change of a sign of the components $a_y$, $a_z$, $b_{xy}$, and
$b_{xz}$. This completes the set of the moment equations derived from the
Fokker-Planck equation \cite{Pokr_art,Pokr_book,Zubarev}, and
gives us the values of $\mathbf{a}$ and $\tensor{b}$.

After $\mathbf{a}$ and $\tensor{b}$ are found, we have for the
perturbations $\mathbf{M}_1$ of the magnetization  and those of
the total stress tensor $\tensor{T}_1^{(2)}$:

$$
\begin{aligned}
M_{1x}=&A_1\gamma_{xx}+A_2H_{1x}^{(2)}\\
M_{1y}=&A_3\gamma_{xy}+A_4\Omega_{xy}+A_5H_{1y}^{(2)}\\
M_{1z}=&A_3\gamma_{xz}+A_4\Omega_{xz}+A_5H_{1z}^{(2)}\\
T_{1xx}^{(2)}=&-p_1+A_6\gamma_{xx}+A_7H_{1x}^{(2)}\\
T_{1xy}^{(2)}=&T_{1yx}^{(2)}=A_8\gamma_{xy}+A_9\Omega_{xy}+A_{10}H_{1y}^{(2)}\\
T_{1xz}^{(2)}=&T_{1zx}^{(2)}=A_8\gamma_{xz}+A_9\Omega_{xz}+A_{10}H_{1z}^{(2)}\\
T_{1yy}^{(2)}=&-p_1+A_{11}\gamma_{xx}+A_{12}\gamma_{yy}+A_{13}H_{1x}^{(2)}\\
T_{1yz}^{(2)}=&T_{1zy}^{(2)}=A_{12}\gamma_{yz}\\
T_{1zz}^{(2)}=&-p_1+A_{11}\gamma_{xx}+A_{12}\gamma_{zz}+A_{13}H_{1x}^{(2)},
\end{aligned}
$$

\noindent where $A_1,\,A_2,\dots ,\, A_{13}$ are known functions
of the applied magnetic field, the frequency of the vibrations, as
well as of the parameters of the ferrofluid, and the equation of
continuity in the form $\gamma_{xx}+\gamma_{yy}+\gamma_{zz}=0$ has
been used. The functions $A_i$ are obtained by means of averaging
(with function $\psi_n(\mathbf{e})$) over the orientations of
chains and averaging over the sizes of the chains (with function
$g_n$).

Thus, the set of governing equation for the first order of the perturbations can be written:

\begin{subequations}
\label{gov_lin}
\begin{align}
\text{div}\, \mathbf{B}_1^{(i)}&=0,\qquad\text{rot}\,  \mathbf{H}_1^{(i)} =0,\quad i=1, 2,  3 \; ,
\label{Maxwell_lin}\\
\text{div}\, \mathbf{u} &= 0 \; ,\label{cont_lin}\\
\rho\frac{\partial u_i}{\partial t}&= \partial_jT_{1ij}^{(2)}\; .
\label{Navier_lin}
\end{align}
\end{subequations}
\noindent Here $\mathbf{B}_1$ is  the perturbation of the induction of the magnetic field.

The linearized boundary conditions differ from those in
Ref.\ \onlinecite{pre02} by the condition for continuation of the stress tensor across the free surface of the fluid

\begin{equation}
\label{bound_lin}
n_j(T_{1ij}^{(1)}-T_{1ij}^{(2)})- \delta_{i,3}\sigma\Delta_\perp\xi =0\quad i=1,\,2,\,3\quad\text{at}\; z=0\, ,
\end{equation}

\noindent where $\Delta_\perp=\partial_{xx}+\partial_{yy}$. In contrast to the previous
case \cite{pre02}, the first two equations in (\ref{bound_lin}) are independent of each
other.

The stability of the flat surface with
respect to standing waves is analyzed by using the Floquet ansatz for the surface
deformations and the $z$-component of the velocity

\begin{subequations}
\label{floquet}
\begin{eqnarray}
\xi(t,x,y)&=&\sin( \mathbf{kr}) e^{(s+i\alpha\omega)t}\sum\limits_{n=-\infty}^{\infty}{\xi_n e^{in\omega
t}} \; ,\qquad\qquad\label{profile}\\
u_z(t,x,y,z)&=&\sin( \mathbf{kr}) e^{(s+i\alpha\omega)t}\sum\limits_{n=-\infty}^{\infty}{w_n(z) e^{in\omega
t}}\label{velocity} \; ,
\end{eqnarray}
\end{subequations}

\noindent where $ \mathbf{k} =(k_x,k_y,0)$ is the wave vector,
$s$ is the growth rate, and $\alpha$ is the parameter determining
the type of the response. For $\alpha=0$ the response is harmonic
whereas for $\alpha=1/2$ it is subharmonic. Expansions similar to
(\ref{velocity}) are made for all other small perturbations and
are inserted into the linearized governing equations
(\ref{gov_lin}). The functions of the vertical coordinate in the
Floquet expansion are of the form $w_n(z)=C_w e^{\pm qz}$, where
$C_w$ is a complex amplitude of the corresponding quantity. The
condition of reality for $\xi(t,x,y)$ leads to the equations
\cite{Kumar}

\begin{subequations}
\label{reality}
\begin{alignat}{2}
\xi_{-n}&=\xi_n^{*},&\quad \alpha&=0 \; , \label{real_har}\\
\xi_{-n}&=\xi_{n-1}^*,& \quad \alpha&=1/2 \; . \label{real_sub}
\end{alignat}
\end{subequations}

Inserting the ansatz into the governing equations (\ref{gov_lin}),
we obtain a set of algebraic equations for the amplitudes of all
perturbed quantities. The solvability condition for this set gives
us four values of the modified wave vector $q$. Hence, the general
solution of the set of governing equations in the ferrofluid
contains eight arbitrary constants. Two more arbitrary constants
are the amplitudes of the perturbations of the field in the air
above and below the fluid after applying the boundary conditions at
$z\rightarrow \pm \infty$. Ten of the boundary
conditions [Eqs. (\ref{bound_lin}) and (3.2a)-(3.2f) in Ref.\
\onlinecite{pre02}] allow one to express all the
perturbed quantities in terms of the coefficients $\xi_n$ which
satisfy the equation

\begin{equation}
\sum\limits_{n=-\infty}^{\infty} {\left( W_{n}\xi_n-a\xi_{n-1}-a\xi_{n+1}\right)e^{[s+i(\alpha+n)\omega ]t}=0} \; ,
\label{disp}
\end{equation}

\noindent where $W_n$ are rather complicated functions of the
applied field, the frequency of vibrations, the depth of the fluid
layer, and of the parameters of the ferrofluid. Here the functions
$W_n$ depend additionally on the coupling constant $\varepsilon$
in contrast to the corresponding functions in the previous paper
\cite{pre02}.

Equation (\ref{disp})  has to be satisfied for all times which
implies that each term of the sum equals to zero. Using the
relations between $\xi_n$ with positive and negative numbers
(\ref{reality}), one gets the set of equations

\begin{subequations}
\label{set}
\begin{align}
W_0\xi_0-a\xi_1^*-a\xi_1&=0,\quad \alpha=0 \; , \label{harm}\\
W_0\xi_0-a\xi_0^*-a\xi_1&=0,\quad \alpha=1/2 \; , \label{subh}\\
W_n\xi_n-a\xi_{n-1}-a\xi_{n+1}&=0,\quad n=1,\cdots \infty \; .\label{any_n}
\end{align}
\end{subequations}

\noindent A cutoff at $n=N$ (in the present work $N=100$) leads to a self-consistent equation for
the acceleration amplitude $a$ \cite{Chen,Chen1},

\begin{equation}
a=|F(a,k,\omega,H_0,\eta,\sigma,\rho,\kappa,\varepsilon)| \; ,
\label{s-const}
\end{equation}

\noindent where $F$ is a complex function expressed in terms of
continued fractions. Equation (\ref{s-const}) can be solved
numerically and gives the dependence of $a$ on $k$ at fixed
parameters. The critical values of the acceleration amplitude
$a_c$ and the wave number $k_c$ correspond to the absolute minimum
of the curve $a(k)$ at zero growth rate ($s=0$).

\section{\label{results} Results and discussion}

\begin{figure}[bp]
    \includegraphics[width=8.5cm]{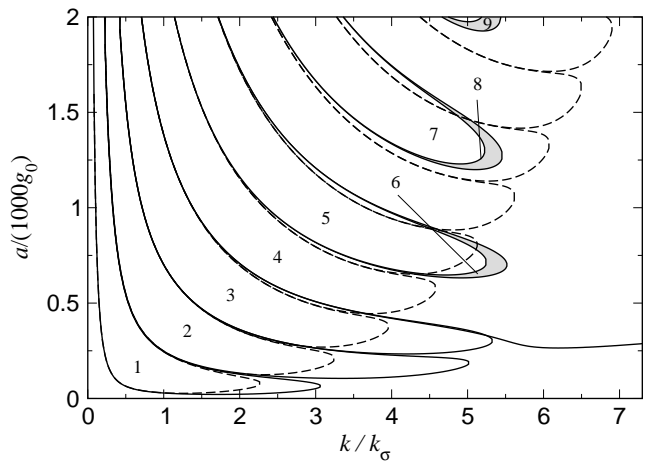}
\caption{Neutral stability curves for the excitation frequency
$f=100$ Hz and layer depth $d=5$ mm. Dashed (solid) lines
are calculated with (without) taking into account deviations of the  magnetization
from equilibrium. The parameters of the fluid are $\eta=0.1$ kg/(ms),
$\varepsilon=5$, $\kappa=0.5$, $\sigma= 0.0265$ N/m, $\rho = 1020$
kg/m$^3$, $R=10.5$ nm, $k_\sigma=614.5$ m$^{-1}$, and the particle volume fraction  $\varphi=0.082$.}
  \label{tongues}
\end{figure}

Figure \ref{tongues} presents marginal stability curves for a
viscous ferrofluid at high frequency for two cases. The dashed
lines are calculated with neglect of the magnetization relaxation
time, $\mathbf{a}=0$ and $\tensor{b}=0$, i.e. the average orientation
of chains immediately follows the field perturbations. The
solid lines depict the neutral stability curves for the system out
of equilibrium. The dependence of the acceleration amplitude  on
the wave number for $s=0$ divides the phase space into regions,
where the surface of the ferrofluid is stable or unstable with
respect to parametrically driven standing waves.
The principal data, which can be extracted, are the critical acceleration
amplitude (scaled with $g_0$), the critical wave number (scaled with the
capillary wave number $k_\sigma=\sqrt{\rho g/\sigma}$),
and the number of the tongue to which
they belong. The number of a tongue $l$ (from left to right) is
the order of response: the basic wave frequency related to the
$l$-th tongue is $\omega_l=l\omega/2$. The odd and even tongues
are the regions, where either a subharmonic or a  harmonic
instability develops. That relation for the different instability types
holds for Newtonian ferrofluids \cite{pre02} but experiences significant
changes if the ferrofluid contains chains.

Due to the finite relaxation time the tongues are deformed. All tongues
but one are now shaped by a lower boundary, a pronounced tip, and a
upper boundary. For the chosen set of parameters in Fig.\ \ref{tongues}
the fourth tongue becomes exceptionally deformed, since it has no tip
and no upper boundary in the way all other tongues have. It is caused by
the fact that the self-consistent equation for the acceleration
amplitude~(\ref{s-const}) has always a solution as $k$ goes to infinity.
To note the difference to all other deformed
tongues, we call it a abnormal tongue because this tongue lacks two essential
features in comparison with all other tongues \cite{note2}.
The typical arrangement of tongues in Fig.\ \ref{tongues} can be generalized as
follows: it is always an even tongue, $l_{\rm abno}=2N$, $N=1,2,\cdots$,
which becomes abnormal (here $l_{\rm abno}=4$). All tongues with
$l>l_\text{abno}$ form separated pairs of two overlapping tongues
(here $l=5, 6$ and $l=7, 8$) inside the abnormal tongue. All
tongues corresponding to a higher order harmonic response, $l=l_{\rm abno}+2L$,
$L=1,2,\cdots$, are "islands" of stability with respect to a response
of the system with the frequency $\omega_\text{abno}$. In
the domains, where the subharmonic tongues overlap with the
regions of the harmonic type of instability (gray regions in
Fig.\ \ref{tongues}), both types of instability can occur.

\begin{figure}[tbp]
\includegraphics[width=85mm]{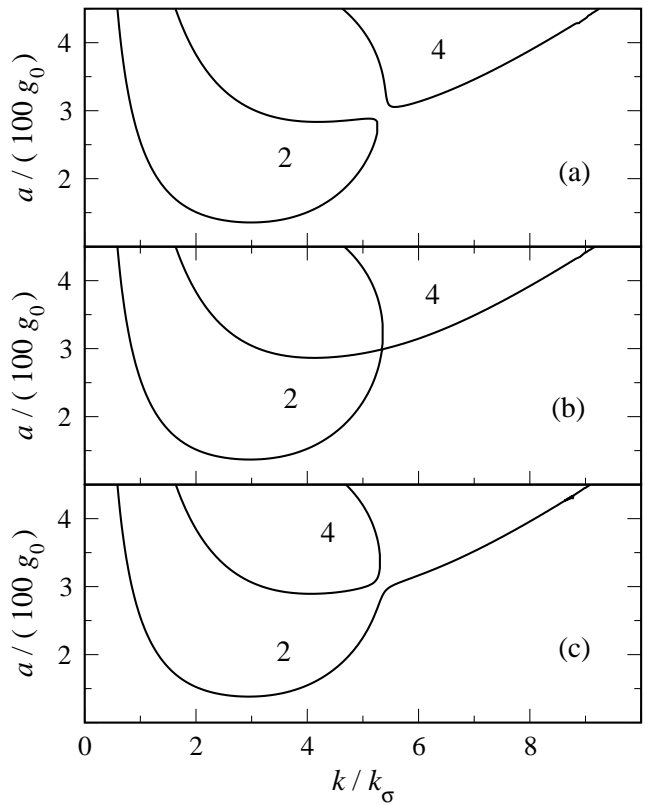}
\caption{The second and the fourth tongues of the neutral stability curves for
the viscosities of the carrier liquid $\eta=0.136$ kg/(ms) (a), $\eta=0.138$ kg/(ms) (b), and $\eta=0.14$ kg/(ms) (c). The remaining parameters are the same as in Fig.\ \ref{tongues}.
} \label{transitions}
\end{figure}

\begin{figure}[tbp]
    \includegraphics[width=8.5cm]{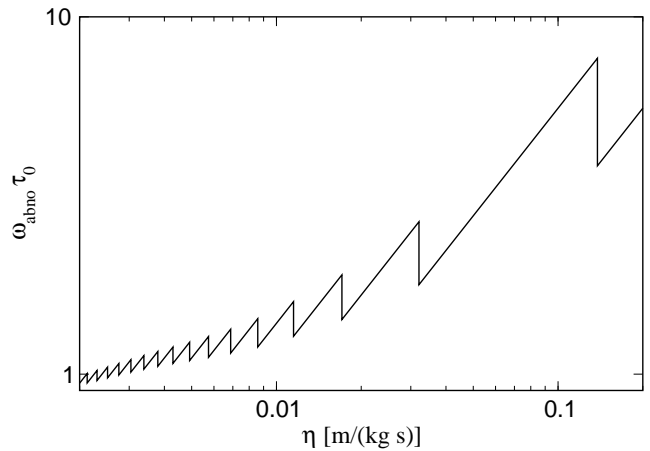}
\caption{Dependence of the product $\omega_\text{abno}\tau_0$ (see text) on the viscosity of the carrier liquid.
The remaining parameters  are
$\varepsilon=5$, $\kappa=0.3$, $\sigma= 0.0265$ N/m, $\rho = 1020$
kg/m$^3$, $R=10.5$ nm and $\varphi=0.082$.
}
  \label{discon}
\end{figure}

Figure \ref{transitions} illustrates the scenario of the transition between
the neutral stability curves with $l_{\rm abno}=4$ and $l_{\rm abno}=2$ with
an increase of the viscosity of the carrier liquid. At the point, where two
tongues touch each other, the amplitudes for the corresponding terms in the
Floquet ansatz (\ref{floquet}) become equal.
The number of the abnormal tongue $l_{\rm abno}$ depends on the mean relaxation
time of a chain $\tau_0 =1/\langle\langle 2D\rangle\rangle$.
The latter can be varied by changing the viscosity of the carrier liquid $\eta$.
In the figure \ref{discon}, the product
$\omega_\text{abno}\tau_0$ is presented as a function of $\eta$.
As the relaxation time increases with the viscosity, the corresponding number $l_{\rm abno}$
decreases: from $l_{\rm abno}=60$ for $\eta =0.001$ kg/(ms) to $l_{\rm abno}=4$
for $\eta =0.1$ kg/(ms). Therefore sequential changes of the number
of the abnormal tongue occur. The corresponding value of
$\omega_\text{abno}\tau_0$ shows a jump-like decrease at those points, where
$l_{\rm abno}$ drops to $l_{\rm abno}-2$.

Both the overlapping of tongues and the appearance of an abnormal
one were also observed for the Maxwell
viscoelastic liquid  \cite{Muller_ve,S_Kumar1}, and it seems that
these features are signs of a viscoelastic behaviour.
One has to note that if the angular distribution of
chains is assumed to be in equilibrium (dashed lines in Fig.\ \ref{tongues}), i.e.
$\tau_0=0$, the neutral stability curves are standard tongues without any
qualitative changes in comparison with the case of a Newtonian ferrofluid
(studied in Ref.\ \onlinecite{pre02}). It shows that it is the combination
of chains {\it and} their nonzero mean relaxation time which
causes the new features in the neutral stability curve.

\begin{figure}[tbp]
    \includegraphics[width=8.5cm]{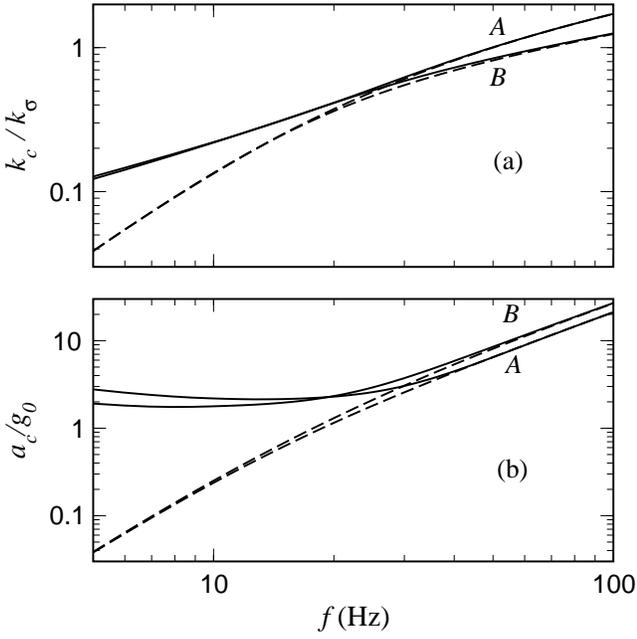}
\caption{Frequency dependencies of the critical wave number $k_c$
(a) and  the critical acceleration amplitude $a_c$ (b) for $d=5$
mm (solid lines) and $d=\infty$ (dashed lines). The magnetization
is assumed to be out of equilibrium for curves $A$ and in equilibrium for curves
$B$. The remaining parameters are the same as in Fig.\ \ref{tongues}.
}
  \label{freq}
\end{figure}

The dependencies of the critical wave number and the critical
acceleration amplitude on the {\em excitation frequency}
$f=\omega/2\pi$ are presented in Fig.\ \ref{freq}. The curves
denoted by $A$ are calculated for a nonequilibrium  magnetization
and the curves $B$ present the case, where the chains follow
the field perturbations immediately. The solid lines correspond to
the ferrofluid layer with depth $d = 5$  mm, and the dashed lines
are calculated for an infinitely deep layer. It is seen that at
high frequencies the critical acceleration in the system out of equilibrium
is lower than that in a model system, where the relaxation
time is neglected. It is clear that with a decrease of the frequency
the deviations of the magnetization from an equilibrium and their
importance become smaller. Therefore, one can expect that curves
{\it A} and {\it B} coincide at low frequencies. However, this is
observed only in the case of the infinitely deep fluid layer.
In the case of $d=5$ mm, the finiteness of the layer and
consequently the influence of the
viscous stresses in the bottom fluid layer become stronger in the
case of the system out of equilibrium, and the critical acceleration
increases more rapidly with a decrease of the frequency. With the
further decrease of the frequency, transitions to higher order
response occur \cite{Cerda,Muller1,pre02}, and the frequency of
the Faraday waves remains high enough to leave the orientation of chains
in nonequilibrium.

\begin{figure}[tbp]
  \includegraphics[width=8.5cm]{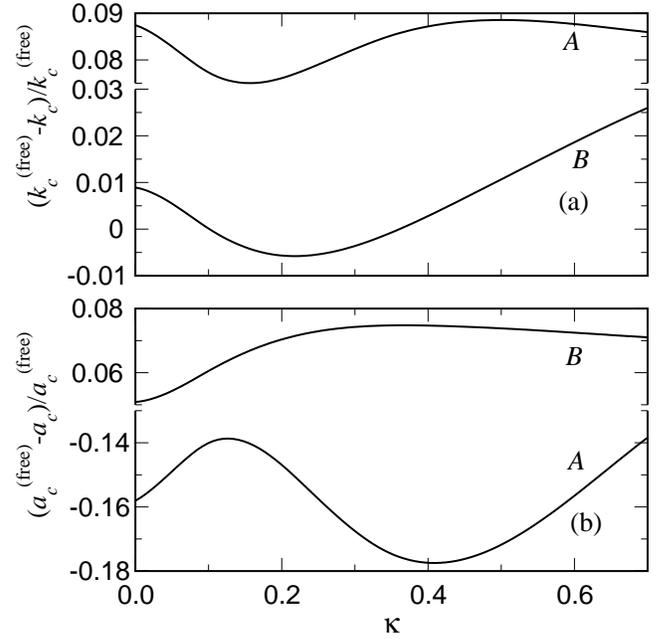}
\caption{The relative differences between the critical wave
numbers (a) and the critical acceleration amplitudes (b) as
functions of the Langevin parameter. The quantities with the
superscript ``(free)'' belong to the chain-free ferrofluid
($g_n=\delta_{n,1}\varphi/v$), the quantities without a
superscript belong to the ferrofluid with chains
[$g_n$ is given by Eq.\ (\ref{sizes})].
Curve $A$: $f=20$ Hz, curve $B$: $f=100$ Hz. The layer depth is $d=2$ mm and  the
remaining parameters are the same as in Fig.\ \ref{tongues}.
}
  \label{field}
\end{figure}

Fig.\ \ref{field} presents the relative differences between the
critical parameters $k_c^\text{(free)}$ and $a_c^\text{(free)}$
for a chain-free ferrofluid and those for a
ferrofluid with chains as functions of the Langevin parameter, i.
e., of the dimensionless magnetic field. The magnetization, chord
$\chi_c= M/H$, and differential susceptibility $\chi_d=\partial
M/\partial H$ of the chain-free ferrofluid are chosen the same as
in the ferrofluid with chains at each point. The magnetization
is out of equilibrium in the ferrofluid with chains and
is in equilibrium in the chain-free one. It is seen, that at $f=20$ Hz
the formation of the chains decrease the critical wave number
[curve $A$ in Fig.\ \ref{field}(a)]. Since the stresses at the
bottom layer are essential at this frequency and $d=2$ mm, this
leads to the increase of the critical acceleration amplitude
[curve $A$ in Fig.\ \ref{field}(b)]. At the higher frequency,
$f=100$ Hz, the changes of $k_c$ are non-monotonic. There is a
range of the parameters, where the critical wave number on the
ferrofluid with chains is larger than that on the chain-free
ferrofluid. The critical acceleration amplitude is decreased by
the formation of chains [curve $B$ in Fig.\ \ref{field}(b)]. Note,
that the susceptibility of the ferrofluid changes strongly with
the increase of the field, therefore the relative importance of
the influences of the magnetic field, the fluid microstructure,
and the viscous stresses depends on the field.

\begin{figure}[tbp]
    \includegraphics[width=8.5cm]{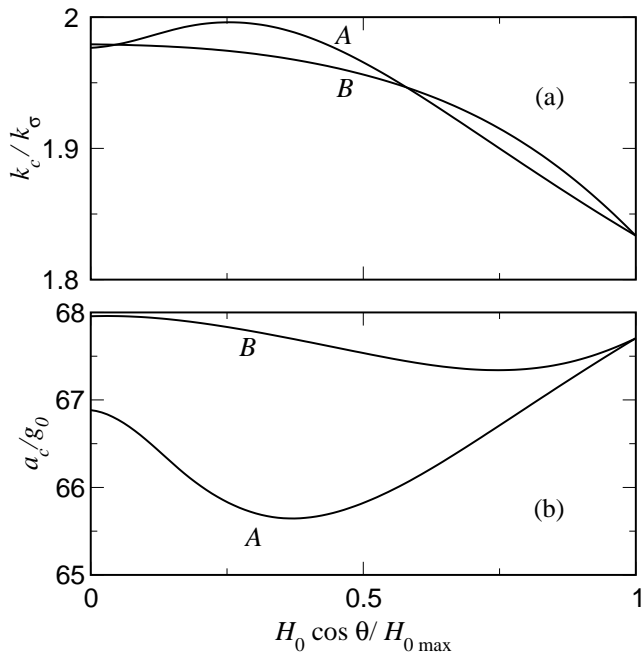}
\caption{The dependencies of the critical wave number (a) and the
critical acceleration amplitude on the projection of
$\mathbf{H}_0$ on the direction of $\mathbf{k}$ for $f=100$ Hz and
    $d=1$ mm. The
curve $A$ is obtained by varying the field strength with
$\theta =0$, whereas the curve $B$ was calculated by varying
$\theta$ with $H_0=H_{0\text{max}}=1.46$ kA/m. Remaining
parameters are the same as in Fig.\ \ref{tongues}. 
}
  \label{angle}
\end{figure}

The above results were obtained for the particular case
$\mathbf{k}||\mathbf{H_0}$. The typical size and orientation of
chains depend on the applied field. Therefore, the threshold of
the instability depends on the strength of the field and the angle
$\theta$ between $\mathbf{k}$ and $\mathbf{H}_0$ separately. It
is not possible to introduce a single parameter like the effective
field \cite{pre02}, which would incorporate those two
dependencies. This fact is illustrated in Fig.\ \ref{angle}. The
product $H_0\cos\theta$ was varied in two ways. For the curves
indicated by $A$, the field strength was changed from  zero to
$H_{0\text{max}}=1.46$ kA/m with $\theta=0$. In the second case
(curves $B$), $\theta$ was changed from zero to $\pi/2$ at the
constant field $H_{0\text{max}}$. In contrast to the chain-free ferrofluid studied in Ref.\
\onlinecite{pre02}, where both curves would be identical, there are now two
different graphs. Whereas the critical wave numbers have similar
values for the two ways of variation, the critical acceleration shows
greater differences: for instance for $H_0\cos\theta/H_{0max}=0.5$ the
difference in
$k_c$ is 0.4 \% against 2.5 \% for $a_c$. Thus it makes a notable difference
for the stability of the free surface whether the system is subjected to
$(\theta =0, H_0=H_{0max}/2)$ or to $(\theta =60^\circ, H_0=H_{0max})$.
This fact suggests a simple method to test the presence of chains in
a sample. If the ferrofluid is chain-free both ways of excitation lead
to the same value of $a_c$, if it contains chains one gets two different
values.

The dependencies are less pronounced than in the previous case,
where $H_{0\text{max}} = 26.6$ kA/m \cite{pre02}, due to the fact
that the range of rather weak fields is studied here. Such a choice of
fields ensures that $\langle n \rangle \le \varepsilon$ and that
the head-to-tail orientation of the moments in a chain is still
preferable which is needed to apply the model.

The nonmonotonic dependence of $a_c(H_0)$ for $\theta=0$
[curve A in Fig.\ \ref{angle}(b)]
caused by the joint action of the two mechanisms of the viscous
damping: i) the dissipation in the bulk fluid and ii) the viscous
stresses in the bottom fluid layer. The first mechanism is
dominant for large wave numbers and causes a decrease of $a_c$.
The second becomes important with the decrease of the wave number
and results in an increase of $a_c$ with field strength. A
detailed discussion of this nonmonotonic behaviour of $a_c(H)$ is
given in Ref.\  \onlinecite{pre02}.

There are three effects caused by the formation of chains. Since a
chain has a magnetic moment larger than that of a single particle,
the formation of chains leads to the increase of the magnetization
of the system at the given field in comparison with a homogeneous
ferrofluid. This strengthens the field influence on the fluid
motion. The effect can be formally taken into account by adjusting
the magnetization and susceptibility of the chain-free ferrofluid
in such a way as it is done in Fig.\ \ref{field}. The two other
effects are more interesting. The chains increase the
magnetization relaxation time and change the effective viscosity
of the suspension both in a steady and in a periodical flow
\cite{Pokr_art,Pokr_book,Zubarev}.

\begin{figure}[tbp]
    \includegraphics[width=8.5cm]{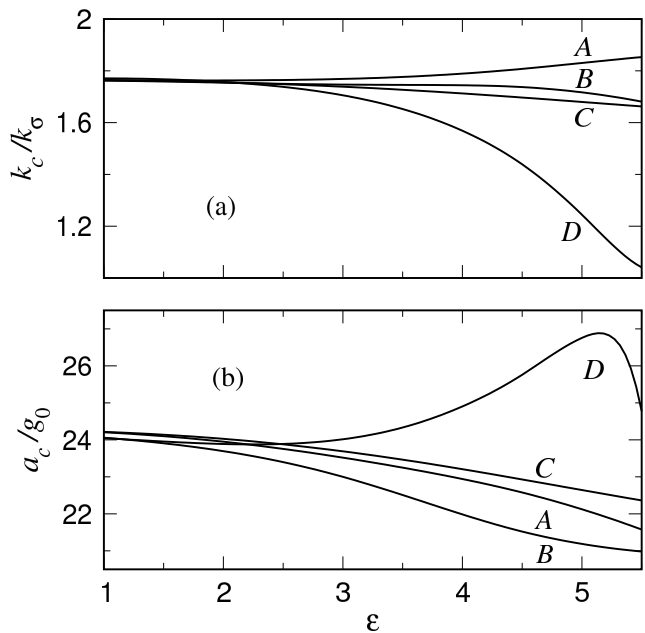}
\caption{The dependencies of the critical wave number (a) and the
critical acceleration amplitude for the infinitely deep layer of
the ferrofluid consisting of $A$: single particles
    with the Brownian
relaxation time $\tau_B=\tau_0$; $B$: chains with a orientations out of equilibrium [Eq.\ (\ref{orient})];
$C$: single particles with zero relaxation time; $D$: chains with
the equilibrium orientations [$\psi_n(\mathbf{e})=\psi_n^0(\mathbf{e})$]. The remaining parameters are the same as
before.
}
  \label{coupling}
\end{figure}

In order to investigate the role of these two effects, a number of
model ferrofluids are considered. The first sample is a ferrofluid
with chains. It is characterized by the mean relaxation time of
a chain $\tau_0$ and unperturbed
values of the magnetization $M_0$
as well as the chord $\chi_c$ and the differential susceptibility
$\chi_d$. The second model ferrofluid contains as well chains,
but in this case the relaxation time is neglected [$\psi_n({\bf e})=\psi_n^0({\bf e})$].
The third and the fourth ferrofluid is chain-free ($g_n=\delta_{n,1}\varphi/v$)
with the same unperturbed values of $M_0$, $\chi_c$, and $\chi_d$.
The magnetization relaxation time for the third ferrofluid is
equal to $\tau_0$, and for the forth sample the relaxation time is
neglected, i.e. the last sample is the ferrofluid, which  was
studied in Ref.\ \onlinecite{pre02}. The predictions of  the
Zubarev model for the case of a chain-free ferrofluid with a
finite relaxation time were compared with the results of a model
using the effective field theory developed by Martsenyuk, Raikher,
and Shliomis in 1974 \cite{MRSH}. In the both cases, the Langevin parameter $\kappa_{adj}$ involved in the model was adjusted to obey $m\varphi/vL(\kappa_{adj})=M_0$,where $L(\kappa)=\coth(\kappa)-1/\kappa$ denotes the Langevin function.
 A fairly good agreement between
the threshold parameters predicted by both  models is observed.
In Figure \ref{coupling} the coupling constant $\varepsilon$ is varied by changing the 
temperature. The intensity of the applied magnetic field $H_0$ is changed in such a way that the Langevin parameter is equal to $\kappa=0.5$.

The comparison of the presented dependencies reveals that the
finiteness of the magnetization relaxation time leads to an
increase of the critical wave number [compare the pairs of the
curves
$A$ -- $C$ and $B$ -- $D$ in Fig.\ \ref{coupling}(a)]. At
the same time, the changes in the slow-flow rheological properties
of the suspension caused by the formation of the chains decrease
strongly the critical wave number (compare curves $C$ -- $D$  and
$A$ -- $B$). The net effect of those two influences gives the
behaviour that is presented by the curves $B$ in Fig.\
\ref{coupling}.

\begin{figure}[tbp]
    \includegraphics[width=8.5cm]{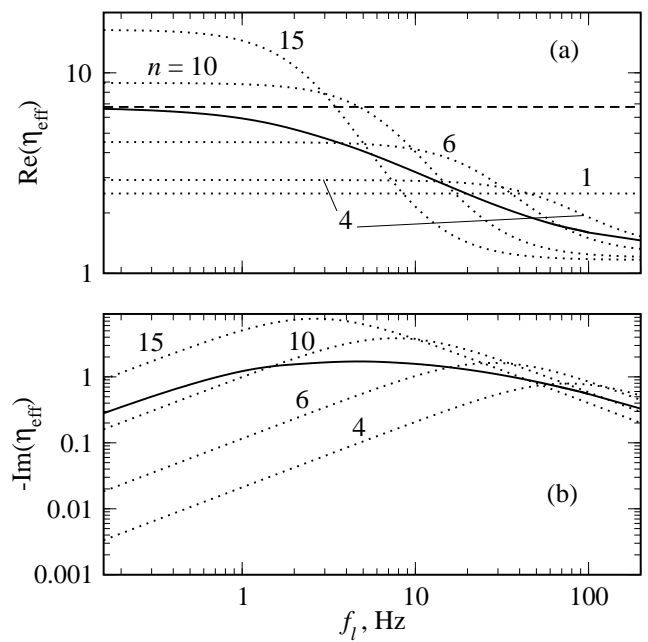}
    \caption{The dependencies of the real (a) and the imaginary (b) part
         of the effective viscosity on the wave
         frequency. The dashed line denotes the  effective viscosity, which
         is calculated with neglect of the relaxation time for
         $\varepsilon=5$.  The solid lines denote the effective viscosity
         calculated using the size distribution of chains (\ref{sizes}),
         and the dotted curves  present the
         real and imaginary parts of the effective viscosity for
         model ferrofluids containing chains of $n$ particles only.
         $H_0=0$, the remaining parameters are the same as before.
}
  \label{ef_visc}
\end{figure}

The critical acceleration amplitude for a ferrofluid with
nonequilibrium chains [curve $B$ in Fig.\ \ref{coupling}(b)] has
the lowest value among all the tested model ferrofluids.
 This can be interpreted as a sign of
an elastic behaviour, which is typical for viscoelastic liquids at
high frequencies. This fact is confirmed in Fig.\ \ref{ef_visc},
where the real and the imaginary parts of the effective viscosity
$\eta_\text{eff}
=\left[\sigma_{ik}/(2\gamma_{ik}\eta\right)-1]/\varphi$ of a
ferrofluid are presented as functions of the frequency of surface
waves $f_l=\omega_l/(2\pi)$ with $l=1$. The case $H_0=0$ is considered,
where the system is
isotropic and the influence of the chains on the rheological
properties of the suspension can be characterized by the single
complex parameter $\eta_\text{eff}$ \cite{Zubarev}. The real part
of the effective viscosity determines the dissipation of the
energy in the system, and a nonzero imaginary part leads to a phase
shift between the tensor of velocity gradients and the stress
tensor, i.e. to a viscoelastic behaviour.
It is seen that the imaginary part of the effective
viscosity is nonzero and has a maximum at a frequency of
$f_l\simeq 5$ Hz [solid line in Fig.\ \ref{ef_visc}(b)],
where therefore a maximal phase shift (viscoelastic
behaviour) can be expected.
 At the same time, the real part of the
effective viscosity decreases [solid line in Fig.\
\ref{ef_visc}(a)]. The effective
viscosity for a stationary flow (with neglect of the relaxation
time) is plotted as dashed line. It is seen that the deviations of
the orientational distribution from equilibrium result in a decrease
of the real part of the effective viscosity. This implies that the
energy is dissipated weaker, and that consequently $a_c$ is lower
than in the case of zero relaxation time.
The comparison of the solid curve with the dotted ones in
Fig.\ \ref{ef_visc}(b) shows that the maximum of the imaginary
part of the effective viscosity is caused by the long chains.

\begin{figure}[tbp]
    \includegraphics[width=8.5cm]{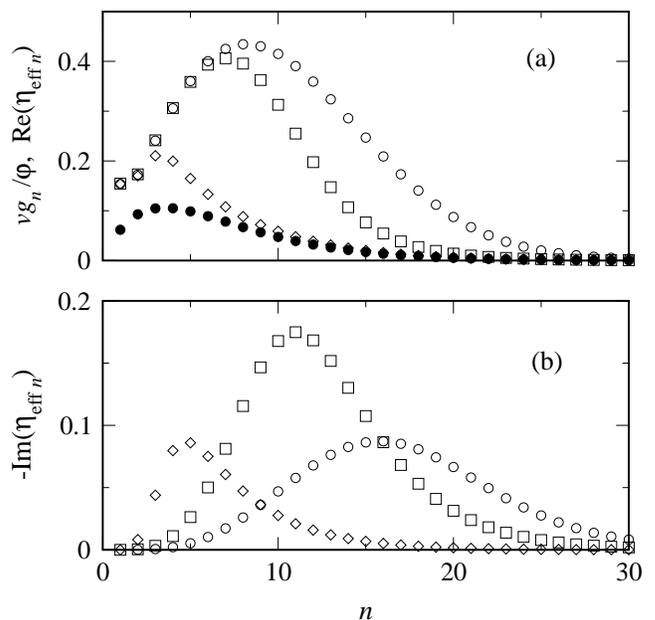}
    \caption{Contributions of chains containing $n$ particles in the real (a)
    and in the imaginary part (b) of the effective viscosity at
    $f_l =1$ Hz ($\circ$), 5 Hz ($\square$), and 100 Hz ($\diamondsuit$).
    The filled circles present the  size distribution of chains (\ref{sizes}).
    The remaining parameters are the same as in Fig.\ \ref{ef_visc}.
}
  \label{impact}
\end{figure}

A comparison of the contributions of chains of different size is done
in Fig.\ \ref{impact} for three values of the wave frequency. The size
distribution of chains (\ref{sizes}) is presented with filled circles
in part (a) of that figure. It is seen that at low frequencies ($f=1$ Hz)
the viscoelastic effect is mainly caused by very long chains with
small values of $g_n$ [see maximum of $-{\rm Im}(\eta_{{\rm eff}\,n})$
at $n=16$ in Fig.\ \ref{impact}(b)].
For $f=5$ Hz the maximum of $-{\rm Im}(\eta_{{\rm eff}\,n})$
is shifted to long chains ($n =11$) and the
value of the maximum increases. Together with the larger values of $g_n$
for those long chains, the total viscoelastic effect is larger than
for $f=1$ Hz [compare the area under the $\circ$- and
$\scriptstyle{\square}$-curve in Fig.\ \ref{impact}(b)].
For high frequencies as $f=100$ Hz the maximum of $-{\rm Im}(\eta_{{\rm eff}\,n})$
is at short chains ($n=5$) and the
value of the maximum decreases. This decrease cannot be compensated by the
even larger values of $g_n$ for those short chains. Thus the total
viscoelastic effect decreases again: the area under the $\scriptstyle{\diamondsuit}$-curve
is smaller than the area under the $\scriptstyle{\square}$-curve.

In summary, by increasing the frequency from $1$ to $100$ Hz the viscoelastic
effect passes at maximum at $f=5$ Hz [solid curve in Fig.\ \ref{ef_visc}(b)]. The
maximal contribution to $-{\rm Im}(\eta_{{\rm eff}\,n})$ comes for $f\lesssim 100$ Hz
from chains off the maximum of $g_n$.
Comparing the area under the $\circ$-, $\scriptstyle{\square}$-, and
$\scriptstyle{\diamondsuit}$-curve in Fig.\ \ref{impact}(a),
one can see that the area is continuously shrinking
with $f$. It results in the solid line in Fig.\ \ref{ef_visc}(a)
describing the weaker dissipation of energy in the system.

The origin of the viscoelasticity (i. e., the dependence of the relation between the tensors of velocity gradients and the stress tensor on the history) in a ferrofluid with chains can be
explain in the following way. When the orientation of a chain is
changed from the equilibrium by the velocity gradients, a restraining moment is
generated by the applied field. Thus the network of chains is an
"elastic" element of the system, which interacts with the viscous
carrier liquid. If the relaxation time of a chain is much shorter than the period of the vibrations, the orientation of chains and consequently the rheological properties of the suspension do not depend on history, and there is no viscoelastic properties. This fact becomes more obvious, if one notes that the relaxation time of a chain is proportional to the carrier liquid viscosity. Thus the negligible relaxation time of a chain implies a weak coupling between the chain and the carrier.

To estimate the importance of the relaxation
processes in a ferrofluid with chains, the Deborah number $De$ is
calculated for the subharmonic response. The Deborah number is the
product of the angular wave frequency with the mean relaxation
time of a chain, $De=\omega_l \tau_0$.
In the case of polymer solutions, the  most pronounced
viscoelastic behaviour should be expected in the range of $De\approx 1$,
whereas at $De\gg 1$ a recovery of the Newtonian behaviour was observed
\cite{S_Kumar}. In the case of ferrofluids with chains, the viscoelastic
behaviour is present over a much wider $De$ range. The frequency range
of Fig.\ \ref{ef_visc} gives $0.033\leq De\leq 3.3$ for which
a viscoelastic behaviour is present since $-{\rm Im}(\eta_{{\rm eff}})
\neq 0$. That fact is supported by the range of appearance for
the abnormal tongues considered to be a characteristic feature of viscoelastic
behaviour. From Fig.\ \ref{discon} one can read that viscoelastic properties
can be expected even for $De$ beyond $10$.

Obviously the viscoelasticity of a ferrofluid with chains differs from that
of a polymer solution. The reason for this difference is the following.
The relaxation time of a long polymer chain involved in the polymer dynamics
is the conformational relaxation time. As soon as the period of excitation becomes
much shorter than the conformational relaxation time, the polymer chain does not participate
in the movement of the solution. Therefore a motion of the Newtonian solvent
through the rigid obstacles of polymer chains is observed. In the case of the rather short
chains formed by magnetic particles, the relaxation time is the orientational
relaxation time of a chain \cite{note1} in the viscous carrier liquid. Since
it is proportional to the viscosity of the carrier liquid, the high values
of $\tau_0$ and $De$, respectively, mean a strong coupling between the motion
of a chain and the carrier liquid, which ensures the viscoelastic behaviour.
Thus ferrofluids widen significantly the range in which viscoelastic
features can be observed in contrast to polymer solutions.

In a real ferrofluid, any variation of the system parameters not only varies
the Deborah number but also affects the critical parameters either directly or
indirectly by changing other properties as permeability or viscosity. In
order to investigate the role of the deviations of the chains from equilibrium
in greater detail, a model ferrofluid with a tunable coefficient of rotational
diffusion for all chains $D_{{\rm tune},n} = D_n D_{\rm scale}$ is studied.
Via that coefficient the mean relaxation time $\tau_0$, i.e. the Deborah number,
can be varied.

$D_{\rm scale}$ is a phenomenological parameter, which is related neither to the
\begin{figure}[bp]
    \includegraphics[width=8.5cm]{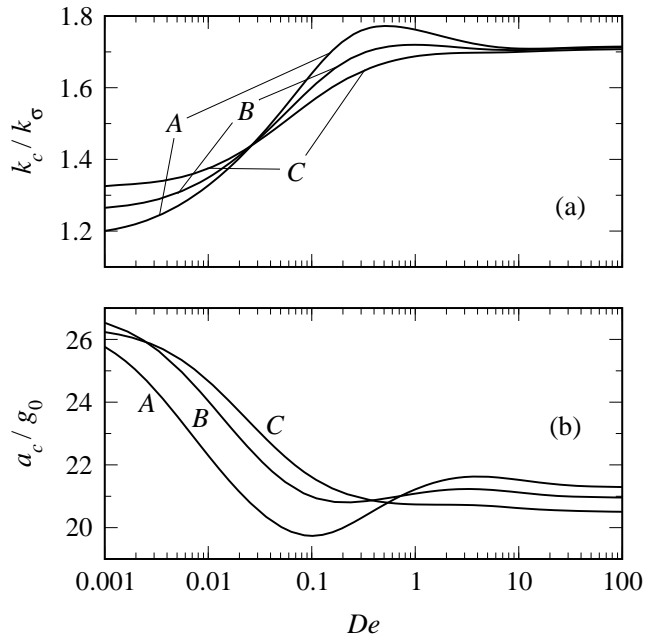}
    \caption{Dependency of the critical wave number (a) and the
    critical acceleration amplitude (b) on the Deborah number for a model
    ferrofluid with the tunable relaxation time of a chain. The
    coupling constant and the Langevin parameter are: Curves $A$:
    $\varepsilon = 5.5$, $\kappa=0.39$,  $B$:  $\varepsilon = 5$, $\kappa=0.5$,
    and $C$: $\varepsilon = 4.5$, $\kappa=0.63$. The remaining parameters are
    the same as before.
}
  \label{Deborah}
\end{figure}
chain size nor to the fluid properties. Using this phenomenological parameter, a
smooth transition can be realized from a model ferrofluid with chains almost in
equilibrium ($D_{\rm scale}\gg 1$ gives $De \ll 1$) to the real ferrofluid
($D_{\rm scale}=1$), and further to a model ferrofluid with chains far from the
equilibrium ($D_{\rm scale} \ll 1$ gives $De\gg 1$).

It is seen in Fig.\ \ref{Deborah} that even small deviations from equilibrium are
significant as the variation of $k_c$ and $a_c$ in the range from $De \sim 0.001$
to $De \sim 0.01$ show. In the case of a system with long chains ($\varepsilon =5.5$),
a pronounced minimum in the dependence of $a_c(De)$ and a maximum in that of
$k_c(De)$ are observed. As $\varepsilon$ and the mean chain length decrease, these extrema
are shifted towards higher values of De and become less pronounced and disappear
for $\varepsilon = 4.5$. Beyond $De\sim O(1)$ one observes a saturation of $k_c$ and
$a_c$ close to the corresponding extremal values. That means that beyond a certain
deviation from equilibrium, viscoelastic effects can neither be enhanced nor
diminished. The latter confirms that there is no recovery to a Newtonian behaviour
at high $De$ numbers.

\section{\label{concl} Conclusion}

The rheological properties of a ferrofluid caused by the formation of
chains is investigated by the linear analysis of the Faraday
instability in a horizontal magnetic field. A horizontally
unbounded ferrofluid layer of a finite depth  has been considered.
The dependencies of the critical acceleration amplitude $a_c$ and
the critical wave vector on the excitation frequency $f$, the
magnetic field ${\bf H}_0$, and the dipolar coupling constant $\varepsilon$ have been
obtained  for different depths of the layer in a wide range of
fluid viscosities. A viscoelastic behaviour is predicted, which is
indicated by the qualitative changes in the neutral stability curve:
an abnormal tongue appears which has no tip and upper boundary in
contrast to all other tongues (Fig.\ \ref{tongues}). Beside its existence it
is revealed how this abnormal tongues is formed by the merging
of other tongues as the viscosity of the carrier liquid is changed
(Fig.\ \ref{transitions}).

The threshold of the instability depends on the applied field
and on the angle $\theta$ between the wave vector
and the applied field separately (Fig.\ \ref{angle}). Therefore
one can easily test a ferrofluid whether it contains chains or
not by choosing different combinations
of $\theta$ and $H_0$ but the same value for $H_0\cos\theta$.
If the ferrofluid contains chains one get different thresholds
whereas the surface of a chain-free ferrofluid is destabilized
at a unique value of $a_c$.

A ferrofluid with chains whose orientation is out of
equilibrium compared with other model ferrofluids (Figs.\
\ref{field}, \ref{coupling}) has the lowest threshold of all.
This is caused by a decrease of the real part of the effective
viscosity [Fig.\ \ref{ef_visc}(a)] which corresponds to a weaker
dissipation of energy and thus to a lower value of $a_c$. The
nonzero imaginary part of the effective viscosity [Fig.\ \ref{ef_visc}(b)]
indicates also the viscoelastic behaviour of the studied ferrofluid.
By analyzing the dependence of the critical parameters on the
Deborah number, it is found that a viscoelastic behaviour occurs
over a much wider range than in viscoelastic polymer solutions.
Particularly no recovery to a Newtonian behaviour at high Deborah
numbers is found.

The present model is restricted to the case of a monodisperse
ferrofluid. An account of the polydispersity will be necessary to
compare the predictions with future experiments using real
ferrofluids. The dipolar interactions between chains and their
flexibility will influence the properties of a real ferrofluid.
Nevertheless, the above-discussed effects should be detectable,
at least qualitatively, in a real experiment.

\begin{acknowledgments}
The authors have much benefited from stimulating discussions with
R. Richter and A. Engel.
This work was supported by the Deutsche Forschungsgemeinschaft under Grant
No. LA 1182/2-3.
\end{acknowledgments}

\begin{thebibliography}{10}
\bibitem{correction}
There is a misprint in \cite{Zubarev} and \cite{Pokr_book}. In the expression
  for $\chi_n$ $\alpha_0'$ has to be in the denominator of the first fraction.
\bibitem{note2}
In Ref.\ \onlinecite{S_Kumar1} the term 'broad harmonic tongue' was used. To
  the authors opinion that term does not note clearly enough the difference to
  all other tongues.
\bibitem{note1}
The estimation for the ratio of the kinematic energy of the chain rotation and
  the thermal energy gives $10^{-8}$ for a chain containing 5 particles and
  $f=100$ Hz, therefore, inertia of a chain is neglected.
\end{thebibliography}

\end{document}